# Electronic Structure and Bonding of Icosahedral Core-Shell Gold-Silver Nanoalloy Clusters Au$_{144-x}$Ag$_x$(SR)$_{60}$

Sami Malola and Hannu Häkkinen*

Departments of Chemistry and Physics, Nanoscience Center

University of Jyväskylä, FI-40014 Jyväskylä, Finland

* Corresponding author: Email: Hannu.J.Hakkinen@jyu.fi, Phone: +358 400 247 973

**Abstract**

Atomically precise thiolate-stabilized gold nanoclusters are currently of interest for many cross-disciplinary applications in chemistry, physics and molecular biology. Very recently, synthesis and electronic properties of "nanoalloy" clusters $Au_{144-x}Ag_x(SR)_{60}$ were reported. Here, density functional theory is used for electronic structure and bonding in $Au_{144-x}Ag_x(SR)_{60}$ based on a structural model of the icosahedral $Au_{144}(SR)_{60}$ that features a 114-atom metal core with 60 symmetry-equivalent surface sites, and a protecting layer of 30 RSAuSR units. In the optimal configuration the 60 surface sites of the core are occupied by silver in $Au_{84}Ag_{60}(SR)_{60}$. Silver enhances the electron shell structure around the Fermi level in the metal core, which predicts a structured absorption spectrum around the onset (about 0.8 eV) of electronic metal-to-metal transitions. The calculations also imply element-dependent absorption edges for Au(5d) → Au(6sp) and Ag(4d) → Ag(5sp) interband transitions in the "plasmonic" region, with their relative intensities controlled by the Ag/Au mixing ratio.

TOC GRAPHICS

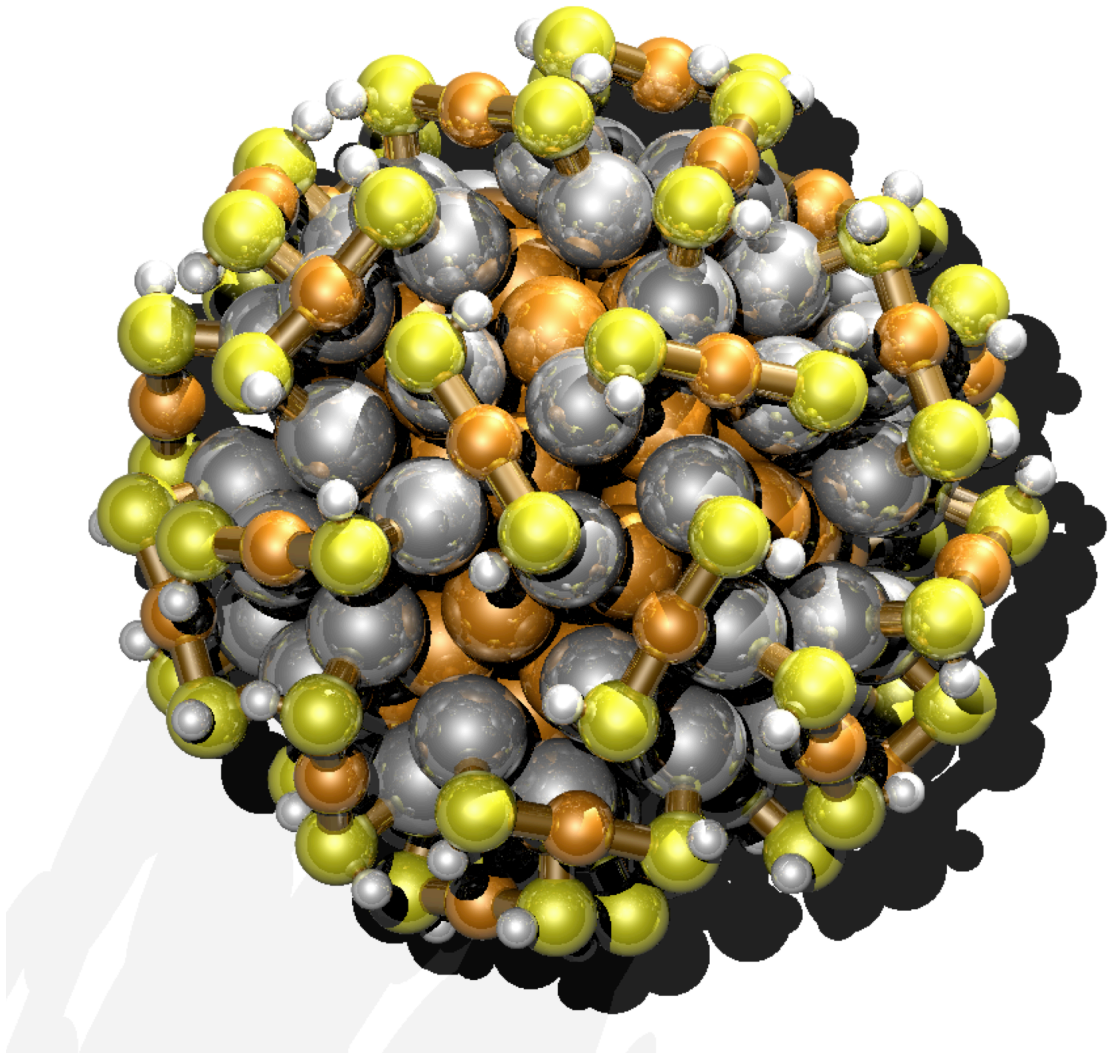

The strong chemical interaction between gold and thiolates (RS⁻) creates ubiquitous Au-S interfaces in thiolate-stabilized gold clusters (nanocrystals),[1,2] colloidal nanoparticles,[3] self-assembled thiolate monolayers on Au(111),[4] and in molecular electronics where organic molecules are frequently linked to gold electrodes via thiolate groups.[5] Since late 1990's, it has been known that one of the most abundant and stable thiolate-stabilized gold nanoclusters has a gold mass of 28-29 kDa (140-150 Au atoms). Its synthesis, electrochemical properties, optical absorption, and bioconjugation properties have been extensively studied.[2,6-17] This cluster is abundantly formed when Au(I)-thiolate complexes are reduced or when colloidal Au(0) particles are etched with excess thiol. It is very stable in ambient conditions and its solubility and chemical derivatisation can be controlled with varying the chemical nature of the thiol (hydrophilic/-phobic). Unlike the smaller clusters $Au_{25}(SR)_{18}^{-1/0}$,[18-20] $Au_{38}(SR)_{24}$,[21] and $Au_{102}(SR)_{44}$,[22] it has so far evaded crystallographic determination of the total structure. Recent high-resolution mass spectrometric experiments have yielded composition assignments for this cluster as $Au_{144}(SR)_{59}$ (ref. 13) $Au_{144}(SR)_{60}$ (ref. 15), and $Au_{144-146}(SR)_{59-60}$ (ref. 16). In 2009, a structural model from density functional theory (DFT) studies for $Au_{144}(SR)_{60}$ was presented, accounting for the reported structural, electrochemical and optical properties of the 28-29 kDa particle.[23] This model features a 114-atom icosahedral metal core that has an inner 54-atom Mackay icosahedron with 60 symmetry-equivalent anti-Mackay sites at its surface, and a protecting layer of 30 RS-Au-SR units. The model follows the general "Divide and Protect" principle[24] where a part of the gold atoms in the cluster (in this case 30 atoms) are covalently bound as Au(I) in the thiolate layer and in a distinctly different chemical state from the 114 core Au(0) atoms. This structural model yields a good agreement with early powder Xray data on the 29 kDa cluster compound[11] and is in close relation to ligand-stabilized $Pd_{145}$ cluster for which the crystal structure is available.[25]

Very recently, Kumara and Dass have shown that co-reduction of gold and silver salts produces nanoalloy clusters with a general formula $Au_{144-x}Ag_x(SR)_{60}$.[26] Three interesting observations were documented: *(i)* the maximal number of Ag atoms mixed with gold seems to saturate near x = 60, achieved with mixing ratio of gold:silver salts by 1:0.66 and 1:0.75, *(ii)* syntheses with higher ratios up to 1:1 did not produce stable $(Au:Ag)_{144}$ clusters, and *(iii)* the optical absorption spectra of the nanoalloy clusters in the UV – vis region exhibit three distinct features: a "plasmonic" peak around 430 nm (2.9 eV) and two shoulders around 310 nm (4.0 eV) and 560 nm (2.2 eV). The higher is x, the more clear is the appearance of these features (the pure $Au_{144}(SR)_{60}$ has a very weak "plasmonic" band at 520 nm). Alloying of gold and silver in nanoscale particles can generally be expected[27] due to the fact that the atom densities in bulk gold and silver are almost identical (the experimental fcc lattice parameters are ca. 4.09 Å for Ag and 4.08 Å for Au) because the significant relativistic contraction of the outermost 6s orbital of Au atom brings its "size" very close to that of Ag). However, the distinct observations by Kumara and Dass clearly call for further theoretical understanding of the alloy clusters $Au_{144-x}Ag_x(SR)_{60}$ particularly when a promising structural model for the all-gold "parent" cluster is available.[23]

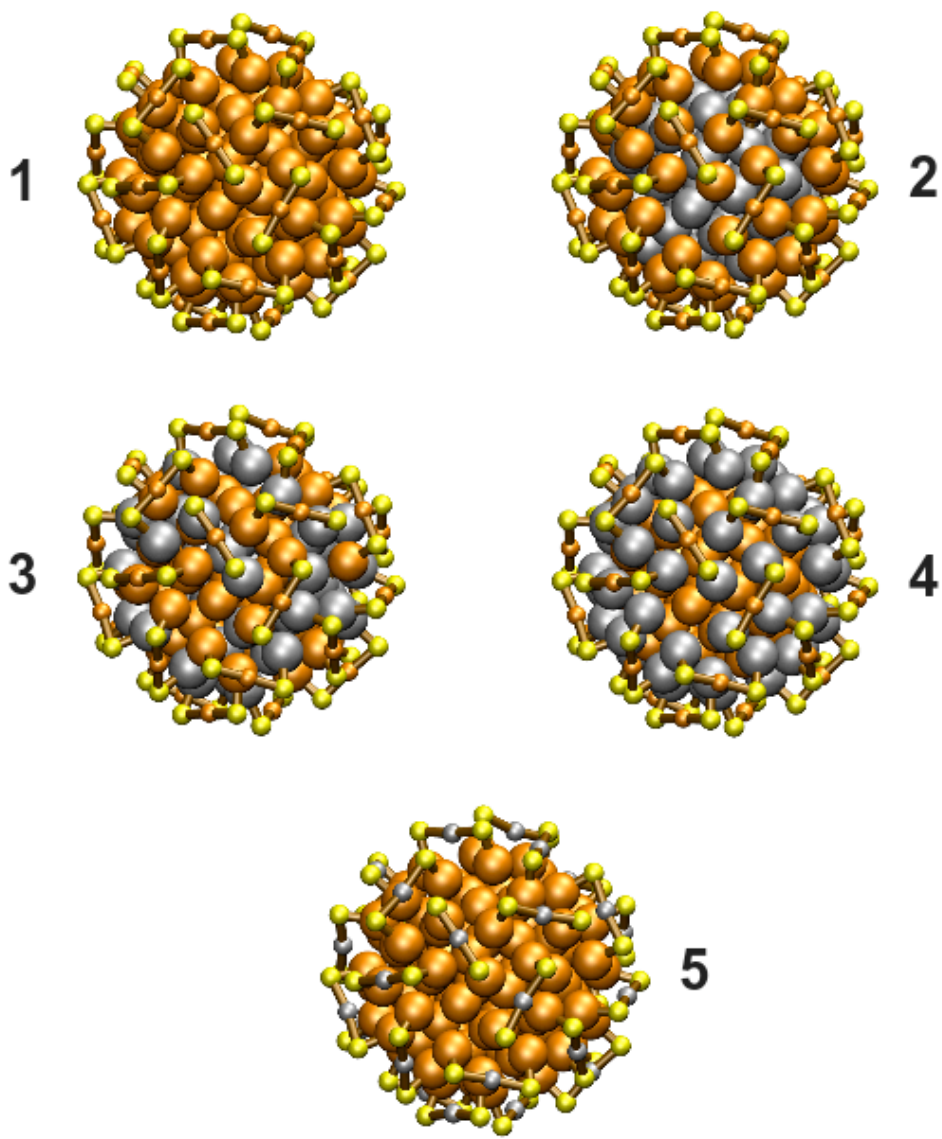

**Figure 1**. Optimized structures of the considered clusters **1 – 5** of composition $Au_{144-x}Ag_x(SH)_{60}$ (see the main text for the detailed descriptions). Gold: brown, silver: shiny grey, sulfur: yellow. The hydrogen atoms are not shown for clarity.

Motivated by this, we employed DFT as implemented in the real-space projector-augmented wave code GPAW[28] (see Computational Details) to study the atomic structure, formation energetics, and electronic structure of a series of $Au_{144-x}Ag_x(SR)_{60}$ clusters with x = 0, 30, 54, and 60, in the context of the previously published[23] structural model for all-gold $Au_{144}(SR)_{60}$. Figure 1 shows the considered structures, labeled as **1** for $Au_{144}(SR)_{60}$ (written as $Au_{114}(RSAuSR)_{30}$ in the "Divide and Protect" scheme), **2** for $Ag_{54}Au_{60}(RSAuSR)_{30}$, where the silver atoms make the inner 54-atom Mackay icosahedron of the metal core, **3** for $Au_{54}Ag_{60}(RSAuSR)_{30}$ where the 60 Ag atoms are randomly distributed in the inner Mackay icosahedron and the anti-Mackay surface layer (30 in both), **4** for $Au_{54}Ag_{60}(RSAuSR)_{30}$ where the 60 anti-Mackay surface sites of the metal core are occupied by silver, and **5** for $Au_{114}(RSAgSR)_{30}$ where silver replaces gold in the thiolate layer. Hence, the motivation to select the used values of x comes from the geometry of the model, and from the experimental observation[26] of the maximum value of x being close to 60. We calculated the formation energies ($E_{form}$) for clusters **1** – **5** and show the results in Figure 2. Based on the definition discussed in Computational Details, the energetically optimal cluster has the lowest (positive) formation energy. Figure 2 shows clearly that $E_{form}$ of cluster **4** has the minimal value, about 0.07 eV per metal atom. This value is half of that calculated for $Au_{144}(SR)_{60}$ (**1**). Clusters where Ag atoms make the inner 54-atom Mackay icosahedron (**2**) or replace Au in the thiolate shell (**5**) are clearly less stable than pure $Au_{144}(SR)_{60}$. Furthermore, the higher formation energy of **3** as compared to **4** indicates that silver and gold prefer separate atom shells inside the core.

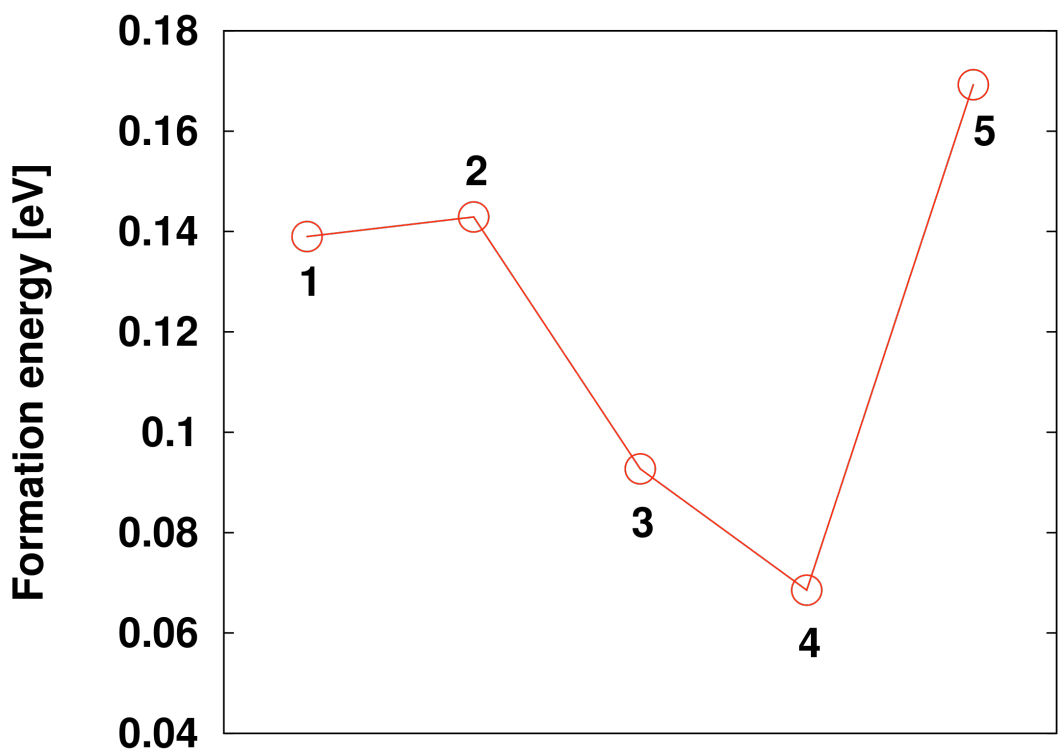

**Figure 2.** The formation energy per metal atom, $E_{form}$, of clusters **1 – 5**. The smallest value indicates optimal formation of cluster **4**.

Confirming the expectations on the grounds of a similar size of Ag and Au atoms, mixing silver into the cluster causes only minor changes in the radial positions of the concentric atom shells (Supporting Information, Figure S1). However, the theoretical XRD functions differ from the all-gold $Au_{144}(SR)_{60}$ due to the different atom form factors of gold and silver, taken here to be linearly dependent on the atomic number (or total electron count).[23] In particular, the XRD function of the energy-optimal custer **4** displays a clear shoulder at $s \approx 3.4$ 1/nm, an enhanced peak at $s \approx 4.3$ 1/nm, and a sharper and inward shifted peak at s ≈ 5.4 1/nm as compared to **1** (Figure S2). These results may prove useful for future experimental analysis of powder samples of $Au_{144-x}Ag_x(SR)_{60}$.

The length distributions of metal-sulfur bonds for clusters **1** and **4** are shown in Figure S3. For cluster **1** the shorter Au-S bonds around 2.34 Å exist in the 30 RSAuSR units and the longer bonds around 2.46 Å connect the sulfurs of the protecting units to the surface gold atoms of the metal core. Related to latter, we observe that for cluster **4** the core-unit Ag-S bonds are considerably longer (ca. 2.52 Å) than the corresponding Au-S bonds (ca. 2.46 Å) in cluster **1**. However the shorter intra-unit RS-Au-SR bonds do not depend on whether the core surface is of gold or silver.

We analyzed the electronic structure of clusters **1 – 5** in the context of the "superatom complex" model.[29,30] This analysis is reported in Figures 3 and S4 (Supporting Information). Figure 3 shows the electron states that are within ± 1 eV from the Fermi level $E_F = 0$ and Figure S4 shows the same analysis for a larger energy range, including also the Au(5d) and Ag(4d) bands. In the neutral state, all these clusters have 144 – 60 = 84 delocalized electron states originating from the Ag(5s) and Au(6s) atomic states. The delocalized electrons organize themselves into energy shells spanning the volume of the metal core, much like electron states in ordinary atoms. In the context of this model, 84 electrons fill a part of the manifold of states with envelope angular momentum symmetries of 1H, 2D, and 3S around the Fermi level (see Figure 3). This manifold spans the states between the closed electron shells at 58 (closing of 1G symmetry) and 92 (opening of 1I symmetry) electrons and is approximately located between -0.7 eV ≤ $E_F$ ≤ 0.15 eV. The result for cluster **1**, in agreement with the one reported previously for $Au_{144}(SCH_3)_{60}$ (ref. 23), shows that 2D and 3S shells are embedded inside the more spread 1H states, which can be understood by considering splitting of the H-symmetry (angular momentum L = 5) by the icosahedral atom (point group) symmetry inside the cluster. The 2D shell remains highly degenerate as expected in icosahedral symmetry. It is seen that the random distribution of 60 Ag atoms inside the metal core of cluster **3** causes maximal disturbance to the shell states including the 2D shell, while the energy-optimal cluster **4** has the set of most degenerate shells in this energy region. This implies a structured low-energy absorption spectrum in the NIR region of electronic transitions. A rather surprising consequence of this symmetric shell structure is that the onset of dipole-allowed electronic transitions is expected at a rather high energy, about 0.8 – 0.9 eV, as indicated by arrows in Figure 3 (from the mixed 2D-1H subshell, located at about -0.35 eV, to the lowest empty 1I subshell located close to 0.5 eV, and from the G shell at -0.9 eV to G shell at the Fermi level). This onset is higher than the one at about 0.5 eV predicted[29] and recently measured[31] for the $Au_{102}(SR)_{44}$ cluster, and is a consequence of the high symmetry of atomic geometry and electron shells of cluster **4**. In the same context, it is intriguing to note that silver enhances the degeneracies of the electron shells of cluster **4** compared to the all-gold parent **1**, where the corresponding transitions are expected at about 0.1 eV lower energies.

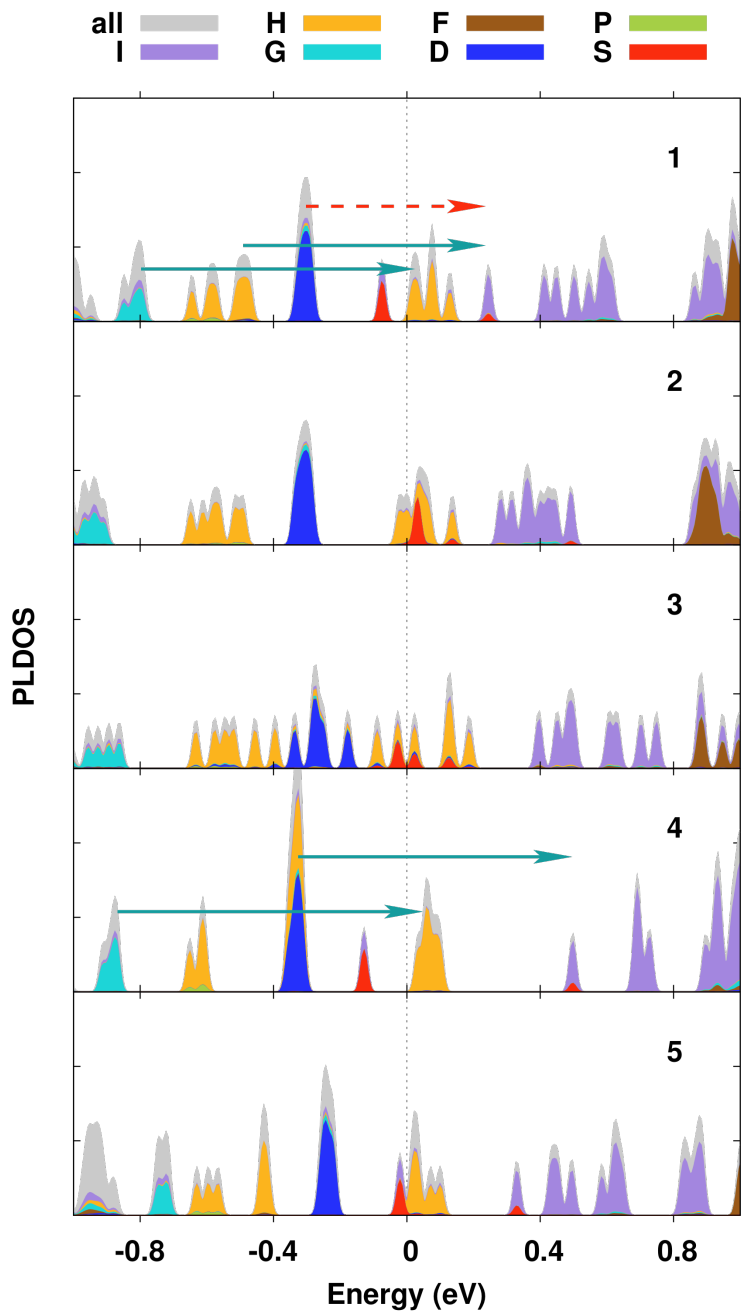

**Figure 3.** Projected local density of states (PLDOS) of clusters **1 – 5.** The projection is done onto spherical harmonics about the cluster center-of-mass as described in ref. 29 selecting the sphere radius as 8.0 Å. This volume contains the 114-atom metal core. We have considered spherical harmonics up to I-symmetry (angular momentum L = 6). The grey area shows the part of DOS that cannot be described in the expansion up to L = 6. The Fermi level is at E = 0 (indicated by the dotted line). The green arrows indicate the lowest dipole-allowed transitions. The transition indicated by red-dashed line is dipole-forbidden.

It is also of interest to analyze the electronic structure by projecting the local density of states into atomic orbitals in different parts of the metal core. Figure 4 displays this analysis for the metal core of the energy-optimal cluster **4**, $Au_{54}Ag_{60}(RSAuSR)_{60}$ that has the inner $Au_{54}$ Mackay icosahedron and 60-atom anti-Mackay silver core surface. Several observations can be made from Figure 4: (i) in the $Au_{54}$ part the Au(5d) band is wide and embedded in the middle of Au(6s) and mixed Au(6sp) states (the states around $E_F$ are dominantly Au(6sp)), (ii) in the $Ag_{60}$ layer the Ag(4d) band is more narrow, totally below the Ag(5sp) states that are dominant around $E_F$, and (iii) the "band edges" for Au(5d) and Ag(4d) are about 1.7 eV and 2.6 eV below $E_F$, respectively (indicated by the black arrows in Figure 4). Thus, it is interesting to note that the expected general differences in the local electronic structures for the gold and silver parts in the core are clearly visible. For colloidal (bulk-like) gold nanoparticles, it is known that the enhanced absorption due to increased joint density of states for 5d → 6sp interband transitions contributes strongly to the surface plasmon (observed for colloidal gold around 520 - 530 nm, i.e., 2.3 – 2.4 eV). For gold nanoclusters larger or equal to $Au_{102}$, a clear increase in absorption takes place around 1.6 – 1.7 eV. [12,31] For silver, the interband transitions for colloidal particles are expected to start around 3.9 eV whence the classical free-electron-type plasmon resonance is already at 3.5 eV.[32] Our calculations offer a first qualitative interpretation of the general features in the absorption spectra measured by Kumara and Dass,[26] predicting that the lowest-energy feature at 560 nm is dominantly due to gold-based transitions in the metal core while the contribution from silver becomes dominant at higher excitation energies. This gives a natural explanation to the fact that the experimentally observed features become more distinct when the silver concentration increases. However, a detailed theoretical analysis and confirmation of this prediction would involve linear-response or true time-dependent DFT calculations of the optical spectrum of cluster **4** for transition energies up to 4 eV. This is a considerable computational challenge for a system of this size, but we are currently working to make this calculation possible. It will be also of great interest to compare such systems to calculations of the smaller nanocluster $Ag_{13}Au_{12}(SH)_{18}^{-1}$ with all-silver core that showed the enhancement of low-energy "superatomic" P $\rightarrow$ D transitions as compared to all-gold $Au_{25}(SR)_{18}^{-1}$.[33]

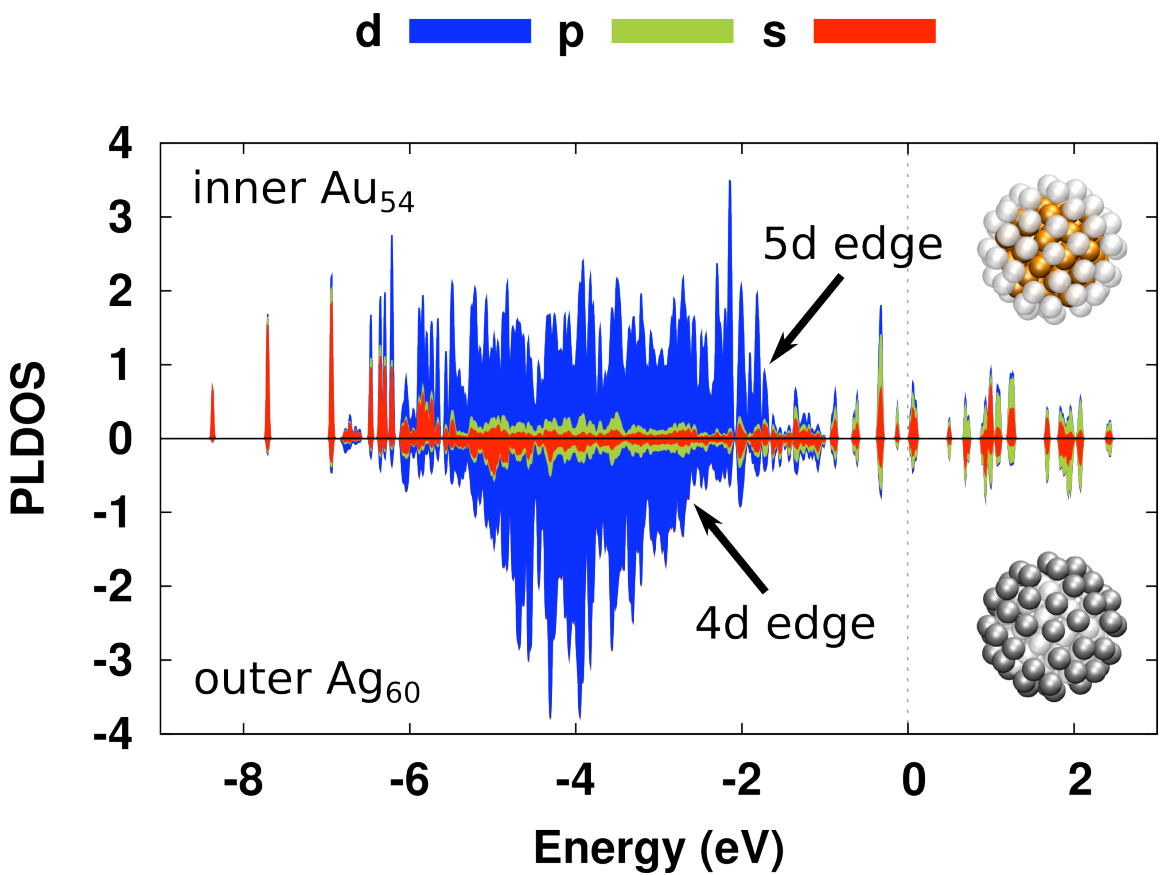

**Figure 4.** PLDOS calculated on atom-centered spheres inside the 114-atom core of cluster **4**, $Au_{54}Ag_{60}(RSAuSR)_{30}$, for the Mackay $Au_{54}$ part (top panel) and for the anti-Mackay $Ag_{60}$ surface layer (bottom panel). The projection is done on atoms as described in ref. 29 selecting the sphere radius as 1.45 Å. The Fermi level is at E = 0 (indicated by the dotted line). The approximate locations of the Au(5d) and Ag(4d) band edges are indicated by arrows.

Perhaps the most interesting result from our calculation is the observation that the "silver-coating" of the metal core is optimal for the cluster formation energy and also increases the degeneracies of electron states, making the electron shell structure more dominant for the properties of $Au_{144-x}Ag_x(SR)_{60}$ clusters as compared to all-gold $Au_{144}(SR)_{60}$. Due to the fact that the reduction/oxidation potentials of gold and silver differ remarkably, it is also of interest to analyze local atomic charges in concentric atom layers of cluster **4** and compare to the all-gold **1**. This analysis is shown in the Supporting Information Table S1. The data shows a very minor excess total negative charge (about -0.7|e|) in the inner $Au_{54}$ icosahedron inside the core of **1**, whence the 60 gold atoms at the core surface and the 30 gold atoms inside the thiolate layer have clearly lost some electron charge to the SH groups (+0.07 and +0.11 unit charges per atom, respectively). At variance with this, the optimal gold-silver cluster **4** has strong variations of charge even inside the metal core: now the $Au_{54}$ inner icosahedron has a total negative charge of -5.7, the 60-atom silver surface has a total charge of +15.0, and the SH groups have a total charge of -11.4. In a way, "silver is reducing gold" in this system in an interesting layer-by-layer fashion. These charge-transfer effects are intrinsic also to the optical properties, but a more detailed theoretical analysis is outside the scope of this letter.

In conclusion, we have reported a detailed DFT calculation and analysis of candidate structures for the 144-metal-atom gold-silver nanoalloy clusters synthesized very recently by the group of Dass. We have found a plausible interpretation to their observation of the maximal silver:gold loading ratio of 60/84, based on our previously published model for all-gold $Au_{144}(SR)_{60}$. Silver is found to preferentially occupy the 60 surface sites of the metal core, and in this configuration silver enhances the electronic shell structure of the particle, that implies structured low-energy NIR absorption spectra around an onset of 0.85 eV and interesting element-dependent "plasmonic" (interband absorption) features in the UV-vis region, controlled by the Ag:Au ratio. One may ask whether the icosahedral geometry would dominate to stabilize larger gold or gold-silver clusters with gold-thiolate surface ligand layer, and whether in the gold-silver alloy clusters the silver always prefers the cover the surface of the metal core. In this context it is intriguing to note that Negishi's group has reported synthesis[34] of stable anionic $Au_{25-x}Ag_x(SR)_{18}^{-1}$ clusters where the maximal x is close to 12, giving a possibility to relate the preferential occupation of silver to the 12 surface sites of the metal core in the known structure of $Au_{25}(SR)_{18}^{-1}$. If this principle is confirmed (from near-future calculations or experiments) also for larger clusters than $Au_{144-x}Ag_x(SR)_{60}$, one has discovered an interesting scheme for manufacturing stable gold-silver nanoclusters with "silver-coated" metal core with potentially interesting and tunable electronic, optical and catalytic properties.

**Computational details**

The structural relaxation, solution to the Kohn-Sham problem, and the electronic structure analysis of $Au_{144-x}Ag_x(SR)_{60}$ clusters were performed using the GPAW code.[28] In this code, the projector augmented wave (PAW) method[35] has been implemented in a real space grid. The $Ag(4d^{10}5s^1)$, $Au(5d^{10}6s^1)$, and $S(3s^23p^4)$ electrons were treated as valence and the inner electrons were included in a frozen core (with the full all-atom electron density available for analysis). The PAW setups for gold and silver include scalar-relativistic corrections. 0.2 Å grid spacing for electron density and 0.05 eV/Å convergence criterion for the residual forces on atoms were used in structure optimizations without any symmetry constraints. The exchange-correlation energy and potential were evaluated using the generalized gradient approximation as derived by Perdew, Burke and Ernzerhof (PBE-functional).[36]

For computational reasons (to minimize the count of valence electrons in the system) we modeled the thiolate with the SH moiety (R=H). The formation energy $E_{form}$ per metal atom for a cluster $Au_{144-x}Ag_x(SH)_{60}$ was defined as

$$E_{form} = \{ E[Au_{144-x}Ag_x(SH)_{60}] + 30\, E(H_2) - (144-x)\, E_{Au,fcc} - x\, E_{Ag,fcc} - 60\, E(SH_2) \} / 144.$$

The respective total energies E[…] of the constituents of the stoichiometric formation reactions were calculated with identical grid parameters in a finite computational cell. The calculations to determine the total energy per atom in *fcc* bulk gold and silver ($E_{Au,fcc}$ = -3.25 eV and $E_{Ag,fcc}$ = -2.75 eV) were done with a 20x20x20 k-point mesh in a periodic cell by using the theoretical equilibrium lattice constants (4.175 Å for both Ag and Au) from GPAW and PBE.

The theoretical XRD intensity was calculated as described in ref. 23. The projection of Kohn-Sham states onto spherical harmonics inside (i) a cluster-centered or (ii) metal atom-centered spherical volume was performed as described in ref. 29. The former analysis (i) yields information of global electron-shell states in the metallic cluster core whence the latter (ii) is an LCAO-type analysis for locations of atomic Ag(4d, 5s, 5p) and Au(5d, 6s, 6p) derived bands. The analysis for local atomic charge (reported in Table S1, Supporting Information) was done by using the Bader method.[37]

**Acknowledgement.** This work is supported by the Academy of Finland and by CSC – the Finnish IT Center for Science. S.M. thanks Olga Lopez-Acevedo for assistance in setting up the initial computation on $Au_{144}(SH)_{60}$.

**Supporting information available**

Radial analysis of atom shells (Figure S1), theoretical powder XRD functions (Figure S2), and full electronic DOS (Figure S4) for clusters **1 – 5**. Metal-sulfur bond-distance analysis (Figure S3) and Bader analysis of local charges (Table S1) for clusters **1** and **4**. This material is available free of charge via the Internet at http://pubs.acs.org.